\pdfoutput=1

\documentclass[12pt,a4paper]{article}

\usepackage{ifthen} 
\newboolean{pdflatex}
\setboolean{pdflatex}{true} 

\newboolean{articletitles}
\setboolean{articletitles}{true} 

\newboolean{uprightparticles}
\setboolean{uprightparticles}{false} 

\newboolean{inbibliography}
\setboolean{inbibliography}{false} 

\usepackage[top=1in, bottom=1.25in, left=1in, right=1in]{geometry}

\renewcommand{\floatpagefraction}{0.99}

\usepackage{microtype}
\usepackage{lineno}  
\usepackage{xspace} 
\usepackage{caption} 

\usepackage{graphicx}  
\usepackage{color}
\usepackage{colortbl}
\graphicspath{{./figs/}} 

\usepackage{amsmath} 
\usepackage{amssymb}
\usepackage{amsfonts}
\usepackage{upgreek} 

\newcommand*\patchAmsMathEnvironmentForLineno[1]{%
\expandafter\let\csname old#1\expandafter\endcsname\csname #1\endcsname
\expandafter\let\csname oldend#1\expandafter\endcsname\csname
end#1\endcsname
 \renewenvironment{#1}%
   {\linenomath\csname old#1\endcsname}%
   {\csname oldend#1\endcsname\endlinenomath}%
}
\newcommand*\patchBothAmsMathEnvironmentsForLineno[1]{%
  \patchAmsMathEnvironmentForLineno{#1}%
  \patchAmsMathEnvironmentForLineno{#1*}%
}
\AtBeginDocument{%
\patchBothAmsMathEnvironmentsForLineno{equation}%
\patchBothAmsMathEnvironmentsForLineno{align}%
\patchBothAmsMathEnvironmentsForLineno{flalign}%
\patchBothAmsMathEnvironmentsForLineno{alignat}%
\patchBothAmsMathEnvironmentsForLineno{gather}%
\patchBothAmsMathEnvironmentsForLineno{multline}%
\patchBothAmsMathEnvironmentsForLineno{eqnarray}%
}

\usepackage[bookmarks=false]{hyperref}    

\usepackage{xspace} 
\usepackage{upgreek}

\def\MagUp {\mbox{\em Mag\kern -0.05em Up}\xspace}

\ifthenelse{\boolean{uprightparticles}}%
{

 \def\PDelta      {\ensuremath{\Delta}\xspace}                 
 \def\PXi      {\ensuremath{\Xi}\xspace}                 
 \def\PLambda      {\ensuremath{\Lambda}\xspace}                 
 \def\PSigma      {\ensuremath{\Sigma}\xspace}                 
 \def\POmega      {\ensuremath{\Omega}\xspace}                 
 \def\PUpsilon      {\ensuremath{\Upsilon}\xspace}                 
 
 \def\PB      {\ensuremath{\mathrm{B}}\xspace}                 
                  
 \def\PD      {\ensuremath{\mathrm{D}}\xspace}

 \def\PK      {\ensuremath{\mathrm{K}}\xspace}

 \def\Pi      {\ensuremath{\mathrm{i}}\xspace}

}
{

 \mathchardef\PDelta="7101
 \mathchardef\PXi="7104
 \mathchardef\PLambda="7103
 \mathchardef\PSigma="7106
 \mathchardef\POmega="710A
 \mathchardef\PUpsilon="7107
                  
 \def\PB      {\ensuremath{B}\xspace}                 
                  
 \def\PD      {\ensuremath{D}\xspace}

 \def\PK      {\ensuremath{K}\xspace}

 \def\Pi      {\ensuremath{i}\xspace}

}

\makeatletter
\ifcase \@ptsize \relax
  \newcommand{\miniscule}{\@setfontsize\miniscule{4}{5}}
\or
  \newcommand{\miniscule}{\@setfontsize\miniscule{5}{6}}
\or
  \newcommand{\miniscule}{\@setfontsize\miniscule{5}{6}}
\fi
\makeatother

\DeclareRobustCommand{\optbar}[1]{\shortstack{{\miniscule (\rule[.5ex]{1.25em}{.18mm})}
  \\ [-.7ex] $#1$}}

  \def\Kbar    {{\kern 0.2em\overline{\kern -0.2em \PK}{}}\xspace}

\def\KorKbar    {\kern 0.18em\optbar{\kern -0.18em K}{}\xspace}

  \def\Dbar    {{\kern 0.2em\overline{\kern -0.2em \PD}{}}\xspace}

\def\DorDbar    {\kern 0.18em\optbar{\kern -0.18em D}{}\xspace}

\def\Bbar    {{\ensuremath{\kern 0.18em\overline{\kern -0.18em \PB}{}}}\xspace}

\def\BorBbar    {\kern 0.18em\optbar{\kern -0.18em B}{}\xspace}

  \def\Y#1S{\ensuremath{\PUpsilon{(#1S)}}\xspace}

\def\Lbar        {{\ensuremath{\kern 0.1em\overline{\kern -0.1em\PLambda}}}\xspace}
\def\LorLbar    {\kern 0.18em\optbar{\kern -0.18em \PLambda}{}\xspace}

\def\AT#1     {\ensuremath{A_{\mathrm{T}}^{#1}}\xspace}           

\def\C#1      {\ensuremath{\mathcal{C}_{#1}}\xspace}                       
\def\Cp#1     {\ensuremath{\mathcal{C}_{#1}^{'}}\xspace}                    
\def\Ceff#1   {\ensuremath{\mathcal{C}_{#1}^{\mathrm{(eff)}}}\xspace}        
\def\Cpeff#1  {\ensuremath{\mathcal{C}_{#1}^{'\mathrm{(eff)}}}\xspace}       
\def\Ope#1    {\ensuremath{\mathcal{O}_{#1}}\xspace}                       
\def\Opep#1   {\ensuremath{\mathcal{O}_{#1}^{'}}\xspace}                    

\newcommand{\tev}{\ifthenelse{\boolean{inbibliography}}{\ensuremath{~T\kern -0.05em eV}}{\ensuremath{\mathrm{\,Te\kern -0.1em V}}}\xspace}
\newcommand{\gev}{\ensuremath{\mathrm{\,Ge\kern -0.1em V}}\xspace}
\newcommand{\mev}{\ensuremath{\mathrm{\,Me\kern -0.1em V}}\xspace}
\newcommand{\kev}{\ensuremath{\mathrm{\,ke\kern -0.1em V}}\xspace}
\newcommand{\ev}{\ensuremath{\mathrm{\,e\kern -0.1em V}}\xspace}
\newcommand{\gevc}{\ensuremath{{\mathrm{\,Ge\kern -0.1em V\!/}c}}\xspace}
\newcommand{\mevc}{\ensuremath{{\mathrm{\,Me\kern -0.1em V\!/}c}}\xspace}
\newcommand{\gevcc}{\ensuremath{{\mathrm{\,Ge\kern -0.1em V\!/}c^2}}\xspace}
\newcommand{\gevgevcccc}{\ensuremath{{\mathrm{\,Ge\kern -0.1em V^2\!/}c^4}}\xspace}
\newcommand{\mevcc}{\ensuremath{{\mathrm{\,Me\kern -0.1em V\!/}c^2}}\xspace}

\def\gsim{{~\raise.15em\hbox{$>$}\kern-.85em
          \lower.35em\hbox{$\sim$}~}\xspace}
\def\lsim{{~\raise.15em\hbox{$<$}\kern-.85em
          \lower.35em\hbox{$\sim$}~}\xspace}

\def\tell1  {TELL1\xspace}
\def\ukl1   {UKL1\xspace}

\usepackage{cite} 
\usepackage{mciteplus}

\begin{document}
\newcommand{\cfft}{C$_{4}$F$_{10}$ }
\newcommand{\cff}{CF$_{4}$ }
\renewcommand{\floatpagefraction}{.8}

\renewcommand{\thefootnote}{\fnsymbol{footnote}}
\setcounter{footnote}{1}


\begin{titlepage}

\vspace*{-1.5cm}

\noindent
\begin{tabular*}{\linewidth}{lc@{\extracolsep{\fill}}r@{\extracolsep{0pt}}}
\ifthenelse{\boolean{pdflatex}}
{\vspace*{-2.7cm}\mbox{\!\!\!\includegraphics[width=.14\textwidth]{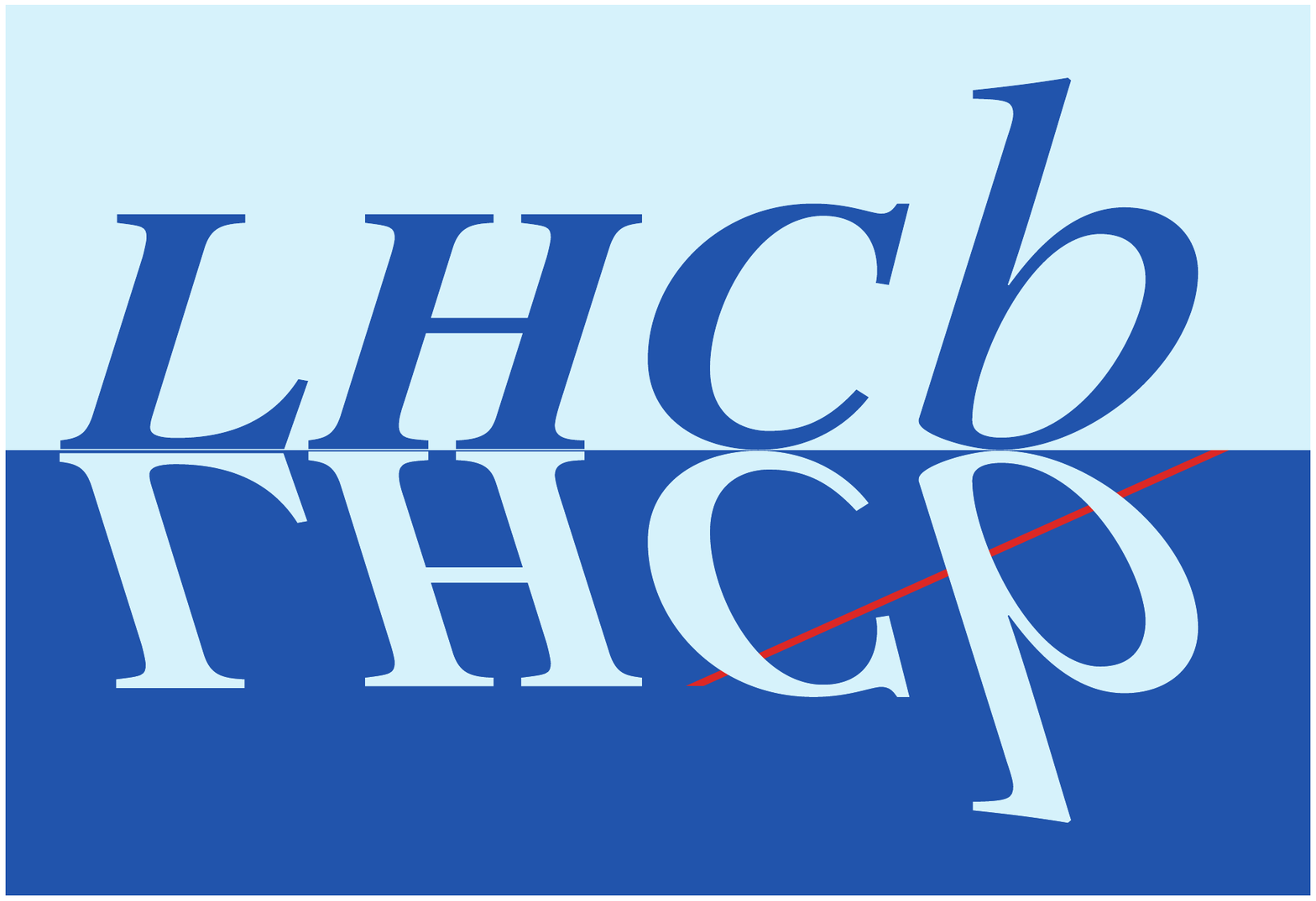}} & &}%
{\vspace*{-1.2cm}\mbox{\!\!\!\includegraphics[width=.12\textwidth]{lhcb-logo.eps}} & &}
 \\
 & & LHCb-PUB-2017-014 \\  
 & & 28 March 2017 \\
 & & \\
\hline
\end{tabular*}

\vspace*{4.0cm}

{\normalfont\bfseries\boldmath\huge
\begin{center}
  The Future of RICH Detectors through the Light of the LHCb RICH
\end{center}
}

\vspace*{2.0cm}

\begin{center}
C.~D'Ambrosio$^1$, R.~Cardinale$^2$, S.~Easo$^3$, A.~Petrolini$^2$, O.~Ullaland$^1$
\bigskip\\
{\normalfont\itshape\footnotesize
$ ^1$European Organization for Nuclear Research (CERN), Geneva, Switzerland\\
$ ^2$Sezione INFN di Genova, Genova, Italy\\
$ ^3$STFC - Rutherford Appleton Laboratory, Didcot, United Kingdom\\
}
\vspace*{0.6cm}
on behalf of the LHCb RICH Collaboration
\end{center}

\vspace{\fill}

\begin{abstract}
  \noindent
The limitations in performance of the present RICH system in the LHCb experiment are given by the natural chromatic dispersion of the gaseous Cherenkov radiator, the aberrations of the optical system and the pixel size of the photon detectors. Moreover, the overall PID performance can be affected by high detector occupancy as the pattern recognition becomes more difficult with high particle multiplicities. This paper shows a way to improve performance by systematically addressing each of the previously mentioned limitations. These ideas are applied in the present and future upgrade phases of the LHCb experiment. Although applied to specific circumstances, they are used as a paradigm on what is achievable in the development and realisation of high precision RICH detectors.
\end{abstract}

\vspace*{2.0cm}
\vspace{\fill}

\end{titlepage}

\pagestyle{empty}  


\newpage
\setcounter{page}{2}
\mbox{~}

\cleardoublepage


\renewcommand{\thefootnote}{\arabic{footnote}}
\setcounter{footnote}{0}



\pagestyle{plain} 
\setcounter{page}{1}
\pagenumbering{arabic}


\section{Introduction}
The two RICH detectors of the LHCb experiment have been operational since 2008 and have been crucial for the physics programme of LHCb~\cite{papanestis}. The current system is expected to continue to take data until 2019, when a two-year shutdown (LS2) is foreseen and LHCb will undergo a significant upgrade. A new RICH system configuration is in preparation (called UPG1), building on the experience gained with the current detector. It is planned to be operational at the start of year 2021~\cite{easo}.

The advent of the High Luminosity LHC from 2027 provides an opportunity for a 10-fold increase in the luminosity at LHCb compared to the present. This would improve the physics reach of the experiment. To meet this challenge, the RICH system needs to be redesigned to improve the Cherenkov angular resolution and to reduce the detector occupancy. In this scenario two new upgrade phases could be foreseen: phase 2a could be introduced in LS3 with a possible further upgrade in LS4 (2b), spanning a period between 2025 and 2035 (UPG2).
 
This paper shows how to overcome the future challenges while providing excellent particle identification within the constraints of the present LHCb experiment. The design draws on recent and future technological advances in photon detection, electronics and optical materials and attempts to systematically reduce all sources of uncertainties. Details will be presented and the expected performance will be reported. 

\section{Occupancies in present and future LHCb}
\begin{figure}
\centering
\includegraphics[width=0.8\linewidth]{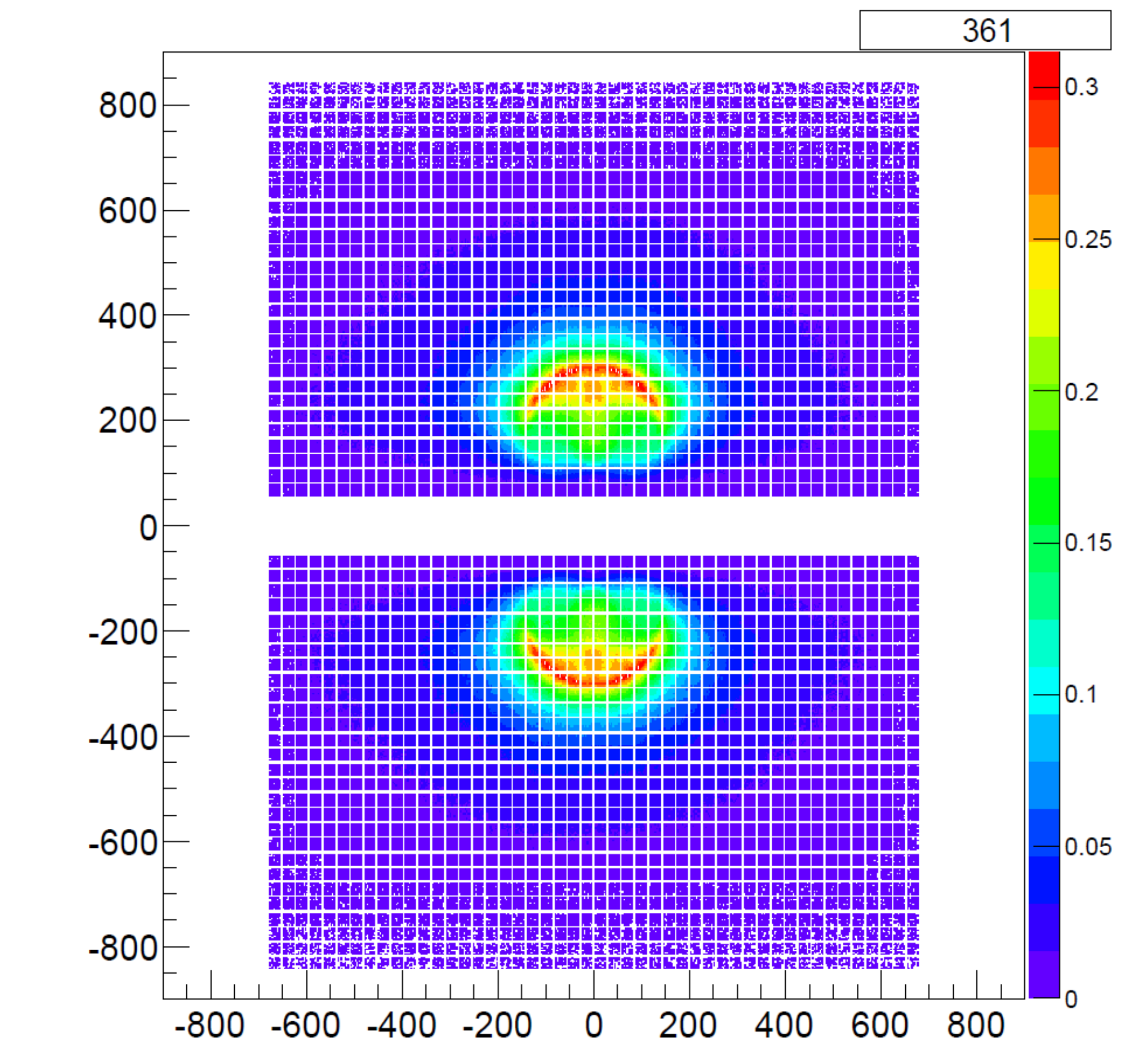}
\caption{Occupancy distribution on the RICH1 photodetector arrays for the UPG1 conditions (location of RICH1 hits on the PMT plane)}. 

\label{fig:occupancy}

\captionof{table}{Corrections applied on pixel size and/or spherical mirror focal length for Upg1 and Upg2, compared to the current conditions.}
\vspace{2mm}
\begin{tabular}{ | c | c | c | c |  }
\hline
             &  Max Peak & Pixel         &  Focal  \\
             & Occupancy & Area          & Length  \\
\hline	
  Current    & 30\%      & 6.3\,mm${}^2$ & 1.35\,m \\
  UPG1      & $\sim 0.5\times 2\times 30\%$ & 6.8\,mm${}^2$ & 1.9\,m \\
  UPG2      & $\frac{1}{14}\times 10\times 30\%$ & 1\,mm${}^2$ & 2\,m \\
\hline  
\end{tabular}
\label{tab:occupancies}

\end{figure}

We start by addressing the occupancy management and take as an example the RICH1 detector alone~\cite{papanestis}. Occupancy is defined as the fraction of detected photons over the total number of channels. At present, the RICH1 detector of LHCb presents peak occupancies of up to 30\% in limited regions of the photon detector array at a luminosity of $\sim\,4\times 10^{32}\mathrm{cm}^{-2}\mathrm{s}^{-1}$~\cite{dambrosio}. The 30\% value is seen as an upper limit for the occupancy in LHCb RICH detectors from simulation results and direct experience, beyond which the particle identification performance starts to deteriorate. In order to operate with luminosities in excess of $\sim\,2\times 10^{33}\mathrm{cm}^{-2}\mathrm{s}^{-1}$, which would double occupancy in the present RICH System upgrade (UPG1), a 2-fold magnification in detection area was implemented  as occupancy scales inversely with the image surface (Fig.\ref{fig:occupancy}). This is achieved by modifying the optics focal length by a factor $\sqrt{2}$ and rearranging mirrors and photodetectors arrays positions~\cite{dambrosio}, (Table~\ref{tab:occupancies}). The small variations in pixel area correspond to the difference between the present (HPDs) and the future UPG1 (MaPMTs) photodetector sizes~\cite{easo}.

 A further jump in luminosity to $10^{34}\mathrm{cm}^{-2}\mathrm{s}^{-1}$ for LHCb would imply an increase in occupancy by a further factor 5. Magnifying the image on the detector plane by the same factor would imply enormous costs and the need for a larger volume inside LHCb. Instead, a reduction of the pixel size (presently $\sim\,$2.6\,mm) to $\sim\,$mm, seems to be realistic, especially when considering the option of covering only the high occupancy regions (Table~\ref{tab:occupancies}). The current readout mode of the LHCb RICH is binary, thus reducing the data size of an event, in order to cope with the high readout rate. However, electronics advances now allow the implementation of a 2-bit readout (conceptually corresponding to a two to three comparators system), which would improve the pattern recognition without affecting significantly the event data size. For further optimisation it can be active only in the high occupancy regions and thus limit the otherwise large increase in channel numbers. In conjunction with an appropriate photon detector, 2-bits logic could further decrease detector occupancy by more than a factor 2.

\section{The role of system uncertainties on pattern recognition and PID performance}
Pattern recognition in high occupancy regions becomes difficult as the overlap between Cherenkov rings of different particles increases (we ignore the tracker performance here and suppose that the trajectory of the particle is exactly known). The detector occupancy can be reduced by using photon detectors with higher granularity, however this is not a sufficient condition to ensure efficient pattern recognition. From basic assumptions, it can be demonstrated that for good pattern recognition $(\sigma_\theta\cdot f)\lesssim\sqrt{A_p}$, where $\sigma_\theta$ is the Cherenkov angle resolution, $f$ is the mirror focal length and $A_p$ is the pixel area. In this scenario, for UPG2 and from Table~\ref{tab:occupancies}, $\sigma_\theta$ should be $\lesssim 0.5\,$mrad, while at present the single-photon Cherenkov angle resolution is $\sigma_\theta\sim 1.6\,$mrad in RICH\,1~\cite{papanestis}.

The reduction in the pixel size, in order to reduce the detector occupancy, has the added benefit of reducing the uncertainty in the Cherenkov angle due to the uncertainty of the photon position measurement. A reduction of a factor 3 would obviously result in a 3-fold smaller pixel uncertainty (Table 1).

\begin{table*}
\centering
\caption{Preliminary simulated performance and photon yields for the various configurations (in bold for UPG2)}
\vspace{2mm}
\begin{tabular}{ | l | c  c  c | c  c | }
\hline
Radiator         & & \cfft &  & \multicolumn{2}{c|}{\cff}  \\
\hline
Detector Version &RICH\,1 & RICH\,1 & \bf{RICH\,1} & RICH\,2 & \bf{RICH\,2} \\
                 &Current (HPD) & UPG1 & \bf{UPG2} & UPG1 & \bf{UPG2}   \\
\hline
Average Photoelectron Yield & 30 & 40 & \bf{60--30} & 22 & \bf{30} \\
\hline
Single Photon Errors (mrad)& \multicolumn {5}{c|}{} \\
\hline
\hspace{5mm}Chromatic      & 0.84 & 0.58 & \bf{0.24--0.12} & 0.31 & \bf{0.1} \\
\hspace{5mm}Pixel          & 0.9  & 0.44 & \bf{0.15}       & 0.20 & \bf{0.07}\\
\hspace{5mm}Emission Point & 0.8  & 0.37 & \bf{0.1}        & 0.27 & \bf{0.05}\\
\hline
\hspace{5mm}Overall        & 1.47 & 0.82 & \bf{0.3--0.2}   & 0.46 & \bf{0.13}\\
\hline  
\end{tabular}
\label{tab:yields}
\end{table*}

Chromatic dispersion affects most the near-UV generated photons, reducing the single-photon Cherenkov angle resolution. The present gases used in the LHCb RICH system, \cfft and \cff, feature low but not negligible chromatic dispersion~\cite{ullaland}. Further improvement through use of a different gas seems difficult. Photon detectors with high quantum efficiency(QE) and a red-shifted spectrum could be employed to circumvent chromatic effects. The high QE is needed to compensate the lower Cherenkov-photon yield at longer wavelengths (green--red) and where the refractive index varies less with wavelength. By convoluting a typical silicon photomultiplier(SiPM) QE curve in the RICH performance simulation, the consequent chromatic error can be reduced by a factor 4 to 8, according to where the wavelength cutoff (from 300nm to 420nm) is applied (Table~\ref{tab:yields}).

Until a few years ago, no existing photon detector seemed to satisfy these requirements. However, the recent progress in silicon photomultiplier technology, their ideal quantum efficiency spectrum, the relatively low cost, the development of cooling techniques to improve dark-count rates and lifetime under severe radiation environments, all seem to provide a robust platform on which to build a vigorous R\&D program to develop efficient and fast single-photon capable devices. Furthermore, they are not affected by magnetic fields, resulting in much simpler, flexible and un-obstructive designs. They will be used in the LHCb upgrade tracker detector and cooled to $\sim\,-40^\circ$C~\cite{fibre}. In that environment, SiPMs will withstand an integrated dose of $\sim\,$100\,Gy in a 10-year operation. This is to be compared to the roughly same dose for RICH2 and to the equivalent of one year of operation in the hottest region of RICH1 (UPG1). Efforts are also being taken up by industry to realise high QE photocathodes in the visible wavelength range.

Finally, emission-point uncertainty contributes to the overall error, as the photons emitted along the particle trajectory arrive at different positions on the photon detector, thus experiencing different optical paths, due to the aberrations of the optical system. The aberrations stem from two main factors: the optical component construction and the angular acceptance (for a mirror, this gives spherical aberration) and the tilts on the components resulting from the necessity to guide the photon outside the acceptance of the experiment (astigmatism). These tilts are often quite large and dictated by the presence of flat mirrors, which have to be placed outside the acceptance (see Fig.2, top), as their material budget (including the supports) would degrade the precision of the tracking detectors. The LHCb RICH-driven development of lightweight composite precision mirrors, allows the placing of flat mirrors within the acceptance, greatly reducing the overall system aberrations. The focal surface becomes close to a plane and the emission-point error is reduced, almost to zero (see Table~\ref{tab:yields}, where the values are simulated following Fig.\ref{fig:layout}, bottom). 

\begin{figure}
\centering
\includegraphics[width=0.8\linewidth]{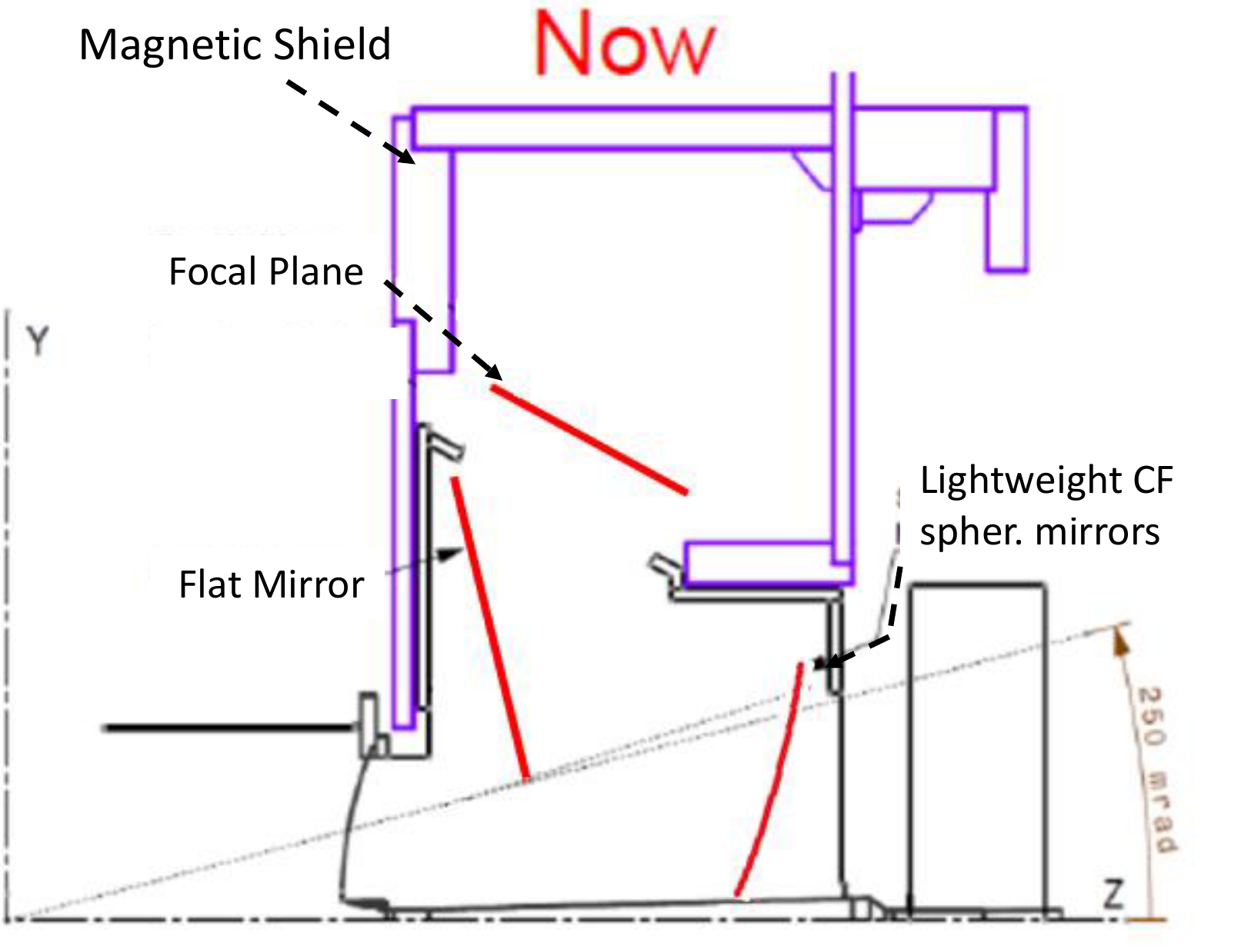}
\includegraphics[width=0.8\linewidth]{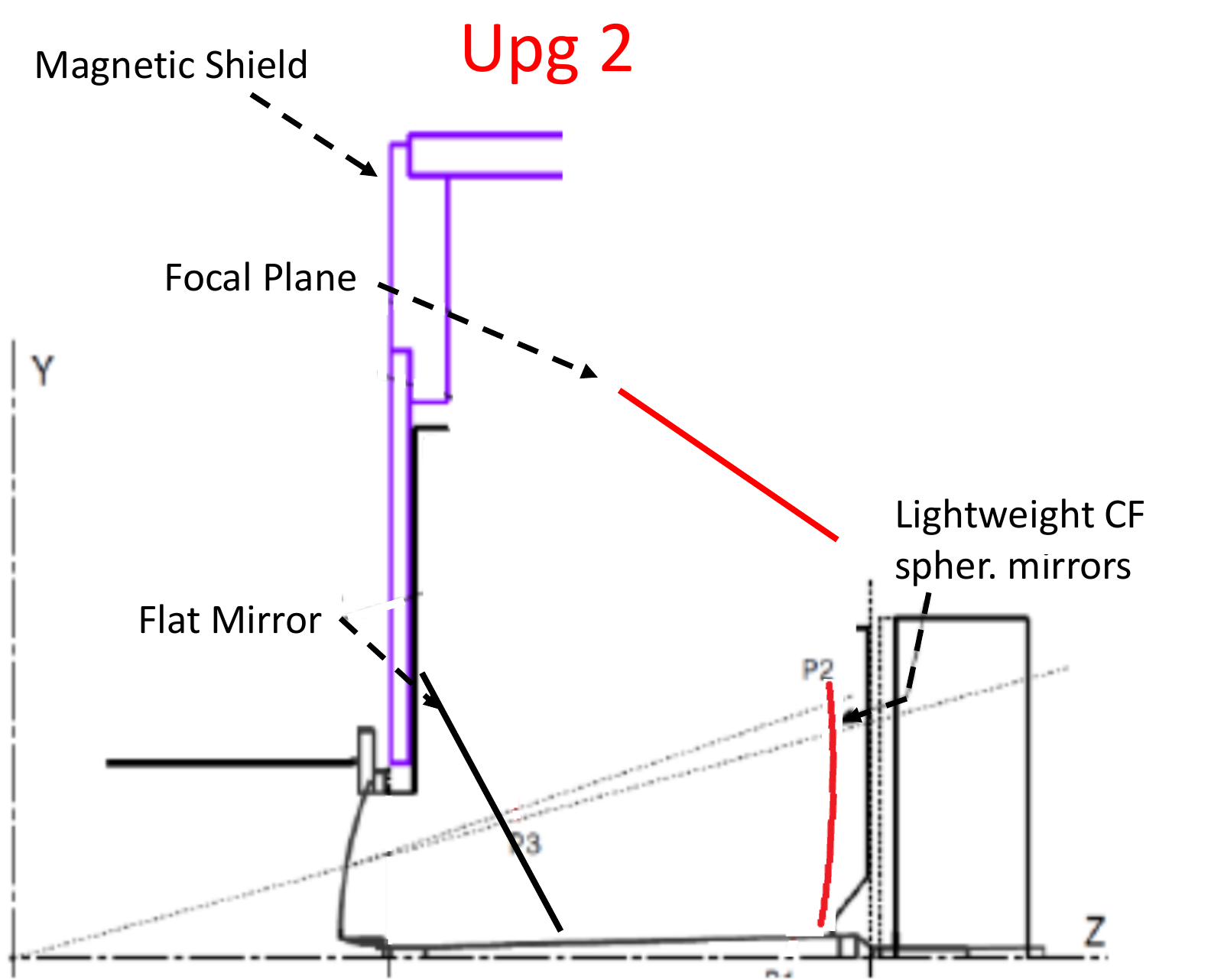}
\caption{(top): Current RICH1 and (bottom): proposed RICH1 UPG2 optical systems.}
\label{fig:layout}
\end{figure}

\begin{figure}
\centering
\includegraphics[width=0.45\linewidth]{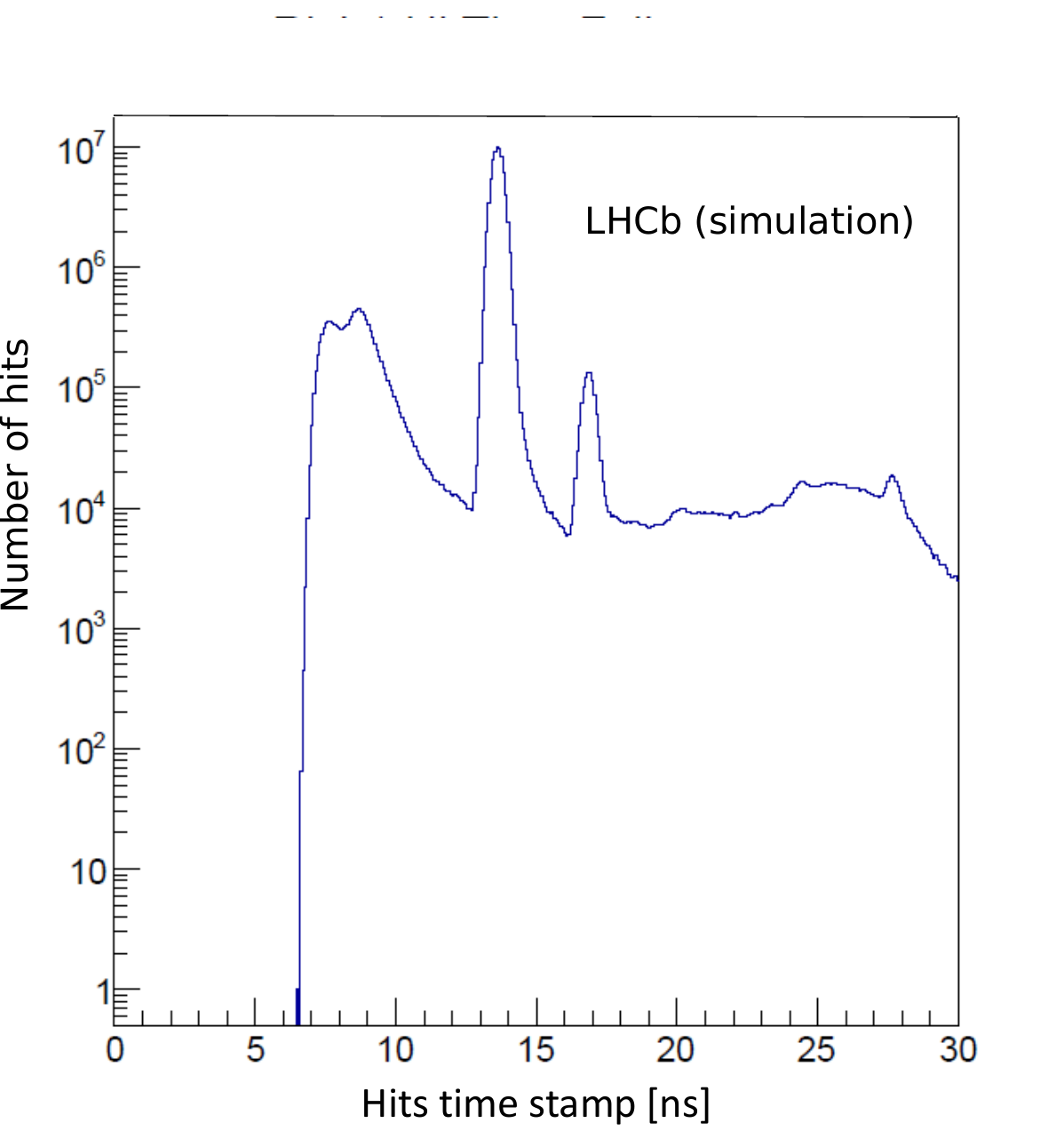}
\includegraphics[width=0.45\linewidth]{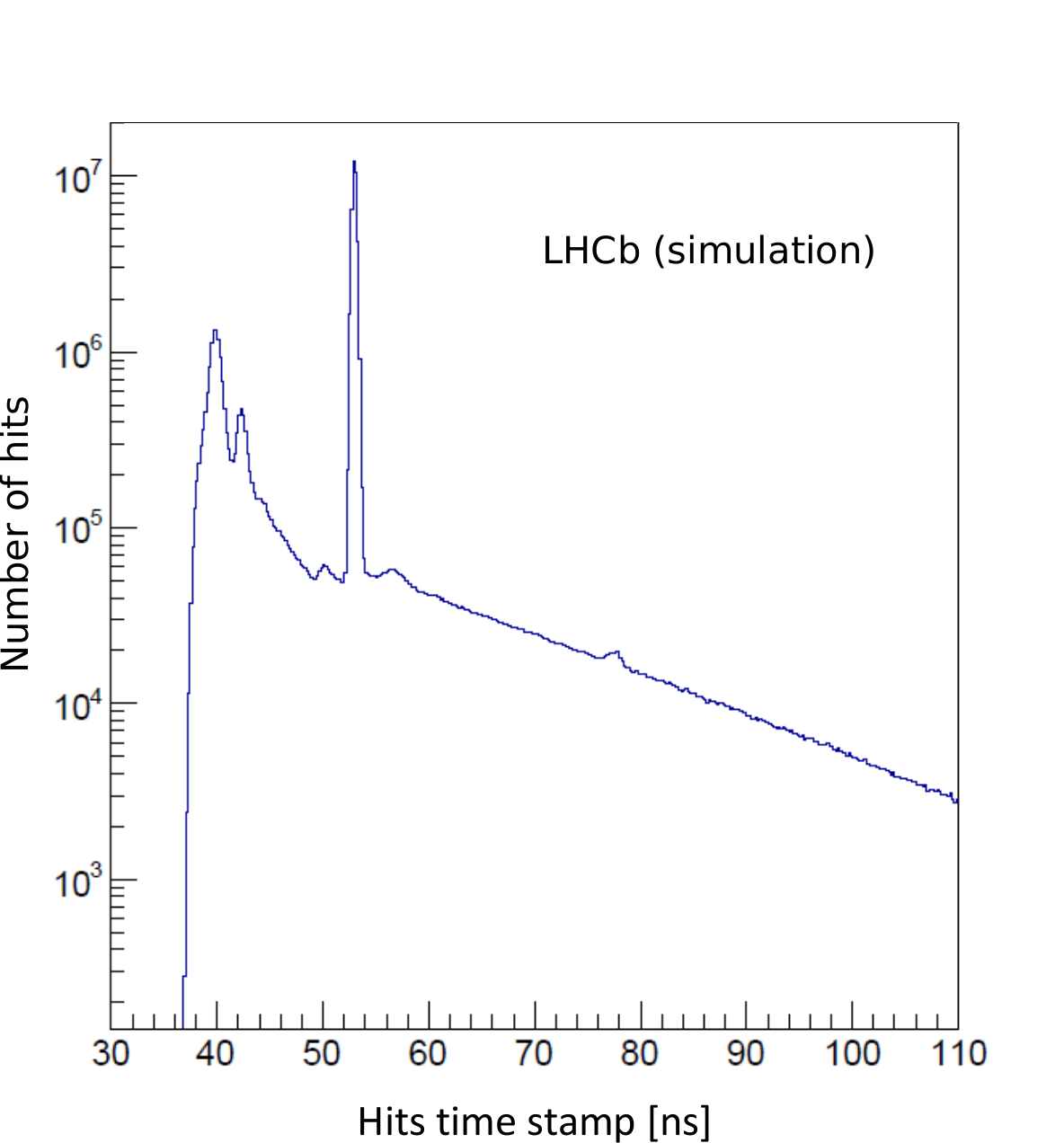}
\caption{Left: Hit arrival time distribution on RICH1 photodetector arrays, for fully simulated events. 
Right: the same for RICH2. The events originate at $z\pm150\,$mm and $t=0\,$s. The main peak is due to hits from particles generating photons(signal), while the first peak(s) in time are due to particles interacting directly with the photodetectors. Other background is attributed to late particles, scintillation, multiple reflections, etc.}
\label{fig:hittime}
\end{figure}

\section{Time resolution for RICH systems}
It is also possible to improve the pattern recognition and to improve the ratio of signal photons to background (electronic noise, dark counts and sources of light other than Cherenkov) by using timing information.

In an aberration- and chromatic-free system, by definition all Cherenkov photons generated by one particle arrive at exactly the same time on the focal plane, regardless of their emission position and wavelength. In reality, time differences between photons generated by the same particle can be kept within picoseconds, even over optical paths of several metres. Therefore, a fast readout system, providing a time stamp of the photon arrival, would require a time resolution of ~1\,ns to deliver a ~150\,ps resolution per ``ring'', resulting from the associated ~40 detected photons (see Table~\ref{tab:yields}).

This is interesting for a few reasons: first, it allows gate durations of ~1\,ns (instead of the present 25\,ns), which would select only true Cherenkov photons from the radiator, resulting from particles produced in the proton-proton collisions. All the other photons, generated by particles directly crossing the photodetectors, curling, spilling over from previous events and generated by radiator scintillation, would arrive outside this $\sim\,$1\,ns window and therefore would be automatically excluded from being acquired (see Figs.\ref{fig:hittime}). Integrating over many bunch crossings, this could result in a reduction of average occupancy by a factor 2. However, the possibility to distinguish particles from different primary vertices, by resolving photons associated to the same ``ring'', would imply a time resolution less than 100\,ps.

\section{Conclusions}

We have described basic ideas for upgrading the LHCb RICH1 detector to operate with luminosities in excess of $10^{34}\mathrm{cm}^{-2}\mathrm{s}^{-1}$, without the need for a complete overhaul of the LHCb detector. The same reasoning is applicable to RICH\,2 (see Table~\ref{tab:yields}) and possibly to other RICH developments.

The ultimate precision in the Cherenkov angle measurement - limited by the performance of the tracking system, particle multiple scattering in the Cherenkov medium and precision of the location of components  - is achieved by finely tuning the optical system, by employing light-weight precision optics, by using low dispersion gaseous media and by shifting the photon detection range to the red. Reducing the detector occupancy requires a compromise between photon detector surface area and granularity and possibly a 2-bit system readout in the “hottest” regions. The system could benefit from the use of precise timing measurements, thanks to fast readout and a quasi-aberration-free optical system.

This is not just wishful thinking. Precision light-weight optics have been and are being further developed and improvements of SiPM-like devices are approaching the maturity and sensitivity required to be eligible for use in a particle physics experiment. We are confident that in ~10 years it will be possible to build such a system. To that end we are going to launch a strong R\&D on photon detector arrays and associated electronics, specific for the LHCb RICH application and in close collaboration with relevant manufacturers. We shall propose this scheme, or its partial implementation, for the possible UPG2a phase to be completed around 2026.




\end{document}